# Chapter 20 Internet of Things

*Edited by Sergio Trilles, Institute of New Imaging Technologies, Universitat Jaume I, Castelló de la Plana, Spain*

E-mail: strilles@uji.es

*Contributors (alphabetically ordered): Carlos Granell[1], Andreas Kamilaris[2], Alexander Kotsev[3]\*, Frank O. Ostermann[4], Sergio Trilles[1]*

[1] Institute of New Imaging Technologies, Universitat Jaume I, Castelló de la Plana, Spain

[2] Department of Computer Science, University of Twente, The Netherlands

[3] European Commission, Joint Research Centre, Italy.

[4] Department of Geo-Information Processing, Faculty of Geographic Information Science and Earth Observation (ITC), University of Twente, Enschede, The Netherlands

\* The views expressed are purely those of the authors and may not in any circumstances be regarded as stating an official position of the European Commission.

**Abstract:**

*Digital Earth was born with the aim of replicating the real world within the digital world. Many efforts have been made to observe and sense the Earth, both from space (remote sensing) and by using in situ sensors. Focusing on the latter, advances in Digital Earth have established vital bridges to exploit these sensors and their networks by taking location as a key element.*

*The current era of connectivity envisions that everything is connected to everything. The concept of the Internet of Things (IoT) emerged as a holistic proposal to enable an ecosystem of varied, heterogeneous networked objects and devices to speak to and interact with each other. To make the IoT ecosystem a reality, it is necessary to understand the electronic components, communication protocols, real-time analysis techniques, and the location of the objects and devices.*

*The IoT ecosystem and the Digital Earth (DE) jointly form interrelated infrastructures for addressing today's pressing issues and complex challenges. In this chapter, we explore the synergies and frictions in establishing an efficient and permanent collaboration between the two infrastructures, in order to adequately address multidisciplinary and increasingly complex real-world problems. Although there are still some pending issues, the identified synergies generate optimism for a true collaboration between the Internet of Things and the Digital Earth.*

*Keywords: Internet of Things, Geospatial standards, Smart scenarios*

## 1. Introduction

According to Jayavardhana (Gubbi et al., 2013), the term Internet of Things (IoT) was first coined by Kevin Ashton in 1999 in the context of supply chain management. Empowered by the latest advances in Information and Communication Technology (ICT), the IoT is revolutionizing the world, opening new possibilities and offering solutions that were unthinkable even only a few years ago. The concept of the IoT is highly multidisciplinary because it brings together a wide variety of technologies, protocols, applications, scenarios, and disciplines (Atzori et al., 2010; Gubbi et al., 2013). The International Telecommunication Union (ITU) Standardisation Sector defines it as *'a global **infrastructure** for the information society, enabling advanced services by interconnecting (physical and virtual) Things based on existing and evolving interoperable information and communication technologies'* (International Telecommunication Union, 2005). As an infrastructure, the IoT can be seen as a broader system involving data, resources, standards and communication protocols as well as theoretical studies.

The pace of IoT development seems quite fast, with continuous proposals of new approaches, applications, and use case scenarios, increasing the presence of IoT in multiple and varied applications, and aspects of daily life. To date, smart devices constitute the IoT's most visible form, applied in a wide range of scenarios and sectors such as cities, industry, commerce, agriculture, home, and mobility. Although we are far from the 200 trillion smart devices as predicted by 2020 (Intel, n.d.), significant progress has been made in this direction. Estimates suggest that there will be 26 smart devices per person in 2020, 40.2% of which will be located in the business environment (termed Industry 4.0).

According to the Forbes analyst Daniel Newman (Newman, 2017), the IoT is one of the most rapidly evolving trends today, especially in three development lines: the analytics arena, the development of edge computing, and the deployment of 5G networks. As 5G technology is progressively implemented and deployed (Shafi et al. 2017), the current analysis platforms will need adaptation in order to analyze effectively the large amount of data flows acquired, produced by IoT devices with increasingly more powerful built-in sensors and emerging real-time analysis functions, empowered even more by the rapid emergence and (parallel) development of edge computing (Shi et al., 2016).

Edge computing is a recent paradigm motivated by bandwidth limitations between the producer (smart objects) and consumer parts (cloud server), as well as the need for improved performance in computing and consumer smart objects. The main feature of edge computing is that data can be processed locally in smart devices rather than being sent to the cloud for further processing.

Like the IoT, Digital Earth (DE) also entails an **infrastructure**. Al Gore, at his famous speech in 1998 (Gore, 1998), introduced the concept of a DE with the vision of extending the real Earth with a digital/virtual replica or counterpart. Over the last two decades, many geographic phenomena and observations have been converted to digital data to be used, analyzed, and visualized using digital tools such as virtual globes (Butler, 2006). In this chapter, we use the term DE to refer to a network infrastructure that allows for the discovery, access, analysis, and processing of spatially referenced data. For more details on DE, we refer the reader to Schade et al. (2019). In particular, Schade et al. describe the origins and evolving concepts of terms such as DE, Geographic Information Infrastructures and Spatial Data Infrastructures, together with their theoretical and technical features. This chapter takes a technological perspective focusing on the description of the current relationships between DE and the IoT, identifying ongoing efforts, potential synergies and bridges, as well as existing limitations and barriers that prevent both infrastructures from collaborating and communicating in practical terms. Instead of operating in parallel, scientists and researchers need the IoT and DE to work jointly by establishing an efficient and permanent collaboration to adequately address the multi-disciplinary nature and growing complexity of the pressing problems that characterize modern science.

The rest of the chapter is divided into five sections. In Section 2, we provide an overview of the most frequent definitions of the IoT, describe our working definitions throughout this chapter, and briefly review related work in the interplay of the IoT and the DE. In Section 3, we analyze the existing interplay between both infrastructures in the context of the main, high-level functions of DE. Then, an overview of relevant case studies across several smart scenarios in which the symbiosis of the IoT and DE could lead to beneficial results is provided in Section 4. Afterwards, Section 5 analyses the frictions and possible synergies today and in the future. Finally, concluding remarks and emerging trends for the immediate future are provided in Section 6.

## 2. Definitions and status quo of the IoT

This section defines the current state of the IoT with respect to the concept of the DE. The first subsection examines the different definitions of a 'Thing', adopted by standardization organizations, followed by our working definition for this chapter. The last subsection describes related works in which interaction between IoT and DE is the main goal.

### *2.1 One concept, many definitions*

The concept of a 'Thing' may seem generic. A 'Thing' can be characterized as a network object or entity that can connect to the Internet directly or through a network gateway. This exemplifies a network-centric perspective of the IoT in which a variety of interrelated 'Things' are able to communicate with each other to deliver new applications and services (Atzori et al., 2010). In contrast to the network-centric vision focusing on the communication technologies being used, the IoT can be seen from a purely Thing-centric perspective in which the services associated with Things are pivotal. These services are expected to manage large amounts of data captured by smart objects or 'Things' as a result of interacting with the environment.

Regardless of the vision, the definition of the term 'Thing' is extensive and includes a wide variety of physical elements. Examples of these elements include: (i) personal objects such as smartphones, smart watches or bands; (ii) ordinary objects and appliances in our daily lives such as refrigerators, lights, cars, and windows; (iii) other identifiable objects equipped with Radio-frequency identification (RFID) tags, Near-field communication (NFC), or Quick Response (QR) codes; and (iv) objects equipped with small microcontrollers.

Because of the heterogeneity of the technology and hardware, there is no single, unified definition of the term 'Thing'. Different international standardization bodies and organizations have suggested a definition, resulting in multiple interpretations of the concepts of Things and the IoT, which sometimes differ only slightly. Consequently, each stakeholder group may have a particular view of what the IoT and Things are, as demonstrated below by the definitions of some internationally renowned organizations.

The World Wide Web Consortium (W3C), an international organization whose aim is the collaborative development of Web standards, defines a 'Thing' as *'the abstraction of a physical or virtual entity that needs to be represented in IoT applications. This entity can be a device, a logical component of a device, a local hardware component, or even a logical entity such as a location (e.g., room or building)'* (Kajimoto et al., 2017).

The Institute of Electrical and Electronics Engineers (IEEE), a global professional engineering organization whose mission is to foster technological innovations and excellence for the benefit of humanity, defines a 'Thing' as a device with programmable capabilities. In contrast to the W3C's definition, the IEEE's definition takes a more practical engineering view of Things, driven by two defining features: (i) Things have the ability to communicate technologically, and (ii) Things have the ability to connect to or integrate in an already connected environment. This networking capability can be based on microcontrollers such as Arduino, Raspberry Pi, BeagleBone and PCDuino, among others.

The European Research Cluster on the Internet of Things (IERC) describes Things as '*physical and virtual things with identities, physical attributes, and virtual personalities and smart user interfaces, and are seamlessly integrated into the information network.*' (IERC, 2014). Similarly, considering that Things belong to a network, the ITU introduces the term *infrastructure* and defines the IoT as *"a global infrastructure for the information society, enabling advanced services by interconnecting (physical and virtual) things based on existing and evolving interoperable information and communication technologies"* (ITU-T, 2012). In addition, the ITU recognizes three interdependent dimensions that characterize Things (Figure 1). This indicates the versatility of the IoT in application domains that differ in terms of the requirements and user needs.

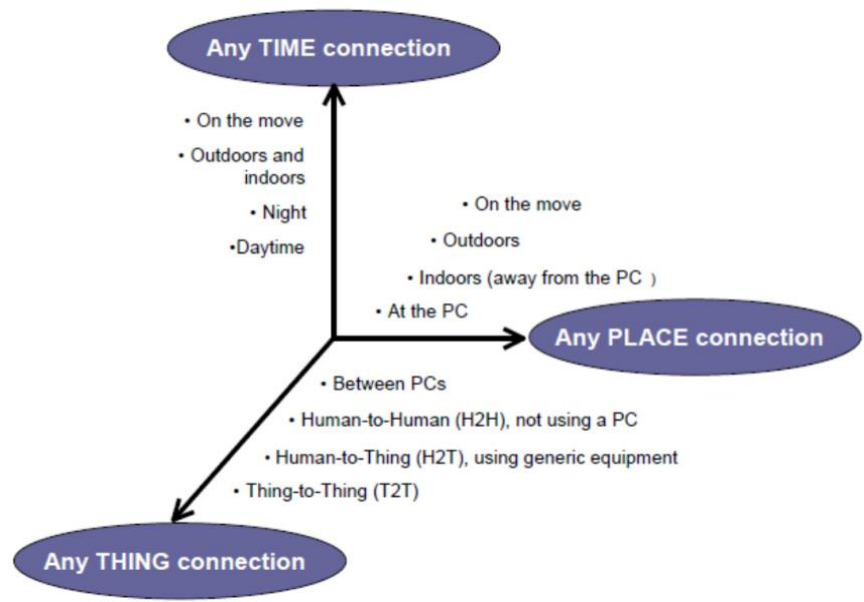

Figure 1: Dimensions of the IoT (Source: ITU, 2012)

The Internet Engineering Task Force (IETF), an open international community of network designers, researchers, and operators concerned with the evolution of the IoT, takes a broad perspective of Things in the context of the IoT, contemplating that *"'things' are very varied such as computers, sensors, people, actuators, refrigerators, TVs, vehicles, mobile phones, clothes, food, medicines, books, etc. These things are classified into three scopes: people, machines (for example, sensor, actuator, etc.) and information (for example, clothes, food, medicine, books, etc.). These 'things' should be identified at least by one unique way of identification for the capability of addressing and communicating with each other and verifying their identities. In here, if the 'thing' is identified, we call it the 'object'"* (Minerva et al., 2015).

Finally, the Organisation for the Advancement of Structured Information Standards (OASIS), a nonprofit consortium that drives the development, convergence and adoption of open standards for the global information society, describes the IoT as a *'system where the Internet is connected to the physical world via ubiquitous sensors.'* (Cosgrove-Sacks, C., 2014). OASIS focuses on the ubiquity of sensors, as they exist in *'every mobile, every auto, every door, every room, every part, on every parts list, every sensor in every device in every bed, chair or bracelet in every home, office, building or hospital room in every city and village on Earth'*.

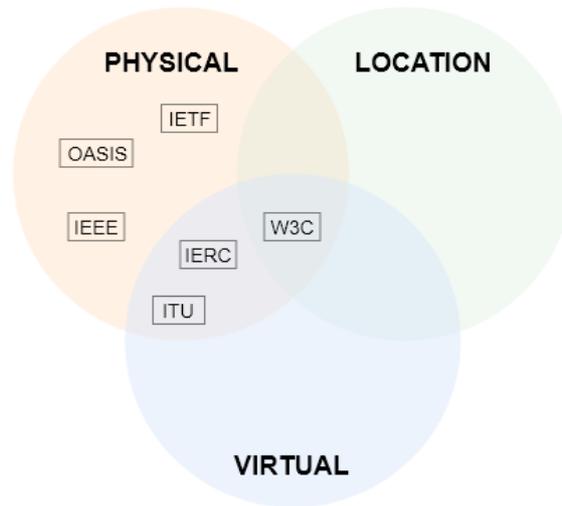

Figure 2: Classification of IoT definitions

Figure 2 categorizes the aforementioned IoT definitions according to a proposed categorization based on physical, virtual and location considerations. The definitions reveal that these institutions and organizations consider the IoT from a physical point of view. In addition to the physical view, three organizations (ITU, IERC and W3C) add a virtual connotation to the definition of a 'Thing'. Only the W3C definition acknowledges explicitly location as a defining element of the IoT.

*2.2 Our definition*

After analyzing the different definitions of internationally renowned institutions and standardization organizations, we propose our interpretation of the term 'Thing' that will be used throughout the rest of the chapter. This definition aims to (i) relate the IoT to DE, and (ii) be as broad as possible.

From our perspective, three main features characterize a 'Thing': (i) networked communication; (ii) programmability (data processing and storage); and (iii) sensing and/or actuating capabilities. From a DE perspective, the third feature plays a more prominent role. The sensing and/or actuating capabilities permit an IoT device or node to interact with its environment. This environment is closely related to the location feature, since all Things will intrinsically have this feature as a property, which increases in importance when the 'Thing' has a mobile component. Contrary to most of the definitions above, we consider a Thing's location as a crucial characteristic because it impacts how a 'Thing' can communicate and how it can interact with its environment. However, we argue that the physical point of view can be understood to include location implicitly, as a physical sensor is located somewhere in the physical world.

*2.3. Early works on the interplay between DE and the IoT*

As noted above, this chapter explores potential bridges between the IoT and DE for the development of applications and services that take advantage of the benefits of both infrastructures to effectively address complex research issues. In this context, we briefly summarize studies related to this objective.

In 1999, Gross predicted that electronic devices would populate the Earth and have the ability to capture different types of information, forming an 'electronic skin' (Gross, 1999). These devices would be able to communicate through the Internet, and include meteorological or pollution sensors, cameras, blood pressure sensors or microphones, among others. The imagined 'electronic skin' could be in contact with what was happening in different scenarios and places on Earth, in the atmosphere, cities, houses, or even in ourselves.

Gross' vision is gradually becoming a reality. There is great variability in the form, size and purpose of sensors in wireless networks. Such wireless sensor networks (WSN) enable distributed communication and data sharing between sensor network nodes. From this perspective, WSN form a subset of the IoT and, as such, the IoT can be seen as the logical next step of WSN in a progression that is still evolving in terms of the sophistication, variability in functionality, flexibility and integration with other infrastructures and network protocols (e.g., the Internet Protocol).

The IoT gained popularity between 2008 and 2013 (Figure 4), and all organizations concerned with WSN began to focus on the IoT. The matured technology of WSN was applied to IoT developments, and DE organizations were not an exception. The field of sensors and sensor networks has been the object of study from multiple and varied angles, including the geospatial community, especially the Open Geospatial Consortium (OGC). The OGC started to transfer improvements made in the definition and application of standards and specifications in the field of WSN to the IoT.

The most significant OGC contribution concerning sensors and WSN has been the Sensor Web Enablement (SWE) standards suite (see Section 2.4 below). SWE enables the discovery and access of sensors and associated observational data through standard protocols and application programming interfaces (API) (Liang et al., 2005, Botts et al., 2008). The SWE standards have been applied directly to many application domains in DE. The shared goal was to observe a particular phenomenon, for example, to predict emergency warnings or fire alarms or alerts when an event is triggered (Wang & Hongyong, 2010). For example, SWE has been widely applied to different Earth Observation (EO) application domains, with disaster management being one of the most important and well-developed. One of the early applications was the use of sensor web techniques to monitor natural and man-made hazards such as fires (Trilles et al., 2014, Jirka et al., 2009, Brakenridge et al., 2003), floods (Brakenridge et al., 2003), and volcanic eruptions (Song et al., 2008).

In parallel with the concept of WSN, Ashton (2009) noted that the term IoT was first used in his work entitled "I made at Procter & Gamble" in 1999. Back then, the IoT was associated with the use of RFID technology. However, the term WSN was not yet the focus of much interest, as shown in Figure 4.

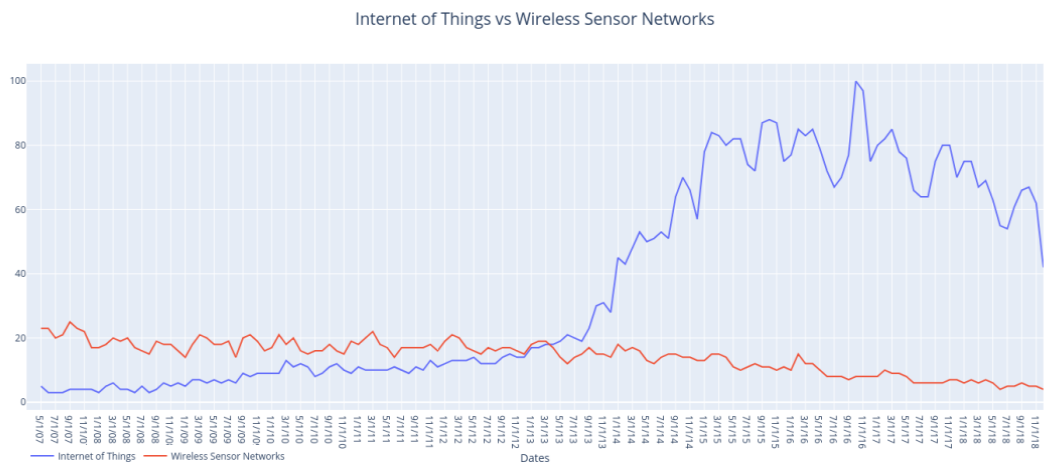

Figure 4: Search volume on wireless sensor networks (red) and the Internet of Things (blue). Source: Google Trends

Some studies explored the connection between the IoT and DE concepts. Li and his colleagues studied the impact of the IoT on DE and analyzed the transition to Smart Earth (Li et al., 2014). The concept was introduced in 2009 during a panel discussion with the U.S. president and U.S. business leaders. In that panel, IBM's CEO Sam Palmisano requested that countries should invest in a new generation of smart infrastructure, with crucial use of sensors, suggesting the concept of 'Smart Earth' as a name. Subsequent governments showed interest in adopting this type of technology, and are making huge investments in researching and developing smart devices (e.g., the 'Array of Things' in Chicago, https://arrayofthings.github.io).

The primary objective of a 'Smart Earth' is to make full use of ICT and the IoT, and apply them in different fields (Bakker & Ritts, 2018). In a 'Smart Earth', IoT devices are placed in all possible locations of our daily life, as long as our privacy can be respected. Through the combination of the IoT, DE, and cloud computing, globally deployed physical objects and sensors can be accessible online. The idea of a 'Smart Earth' is ambitious and includes remote sensing, GIS and network technology in combination with DE platforms (see Mahdavi-Amiri et al., Chapter 2 in this book featuring "Digital Earth Platforms"). The goal is to enable sustainable social development, which is a visionary step that is still utopian today, towards the establishment of a global information infrastructure to support

UN Sustainable Development Goals (see Clinton et al., in this book, Chapter 12 "Digital Earth for UN SDGs").

The work by van der Zee & Scholten (Zee & Scholten, 2014) highlighted the importance of location in the concept of the IoT. The authors noted that space and time can play a role as 'glue', to enable an efficient connection between smart devices; therefore, geospatial sciences should have an active presence in the development of IoT architecture. In their study, van der Zee & Scholten described a set of technologies related to the geospatial domain and big data analysis that could be combined with the IoT. The authors concluded that these technologies were already available for application in the field of the IoT and recommended their immediate use. However, the authors also identified the lack of IT professionals with knowledge in geospatial sciences as the main obstacle in massive uptake of the IoT for geo-related applications. They proposed to address this limitation through a gradual incorporation of core geospatial skills and competences into IT curricula.

Our aim in this chapter is to move beyond the initial steps and thoughts presented in van der Zee & Scholten (2014), where the status quo of the IoT and DE was described five years ago. We focus on the 'current status quo' by outlining emerging technology trends that can be crucial for establishing real connections between DE and the IoT, and investigate developments during the last five years in particular. Even though development has been gradual and incremental, and not rapid and revolutionary (i.e. from a GIScience perspective), new requirements and technology trends have appeared and the IoT has become a topic that is undoubtedly gaining increasing traction.

*2.4. IoT standards initiatives from DE*

As noted above, the IoT ecosystem has been very diverse for several years (Atzori et al., 2010), and its diversity has been increasing. It is comprised of heterogeneous devices, protocols and architectural approaches. A plethora of international initiatives are put in place to unify and streamline aspects associated with the design and implementation of IoT infrastructures. The current standardization initiatives address aspects related to discoverability, data transmission, device processing and tasking.

The growing number of interconnected devices, combined with the increasing importance of the use of the IoT in almost any aspect of human life, tend to increase the need and importance of mature, well-established and -implemented standards. The diversity of different standardization initiatives provides designers and developers with a broad range of opportunities that do not necessarily complement each other. There are multiple ways of reaching the same destination, i.e., there is no single solution to be adopted. Here, we provide a short overview of selected IoT standards that play an important role within the context of DE. The SWE suite of standards is described in more detail in Chapter 8 of this book.

From the geospatial perspective, the OGC coordinates different standardization initiatives. This consortium is comprised of more than 525-member organizations from governmental, commercial, non-governmental, academic and research institutions. The primary objective of the OGC is to develop open standards that include a geospatial component. These standards are developed through a consensus-based process and are openly available to streamline the exchange of geospatial data. OGC standards are used in a wide variety of domains, including geosciences and the environment, defense and intelligence, emergency and disaster management, and public services, among others.

Over a decade ago, well before the IoT became mainstream, the OGC developed the SWE suite of standards for spatio-temporal observation data (Botts et al., 2008). SWE outlines a set of specifications related to sensors and proposes data models and Web service interfaces that can act as a bridge between sensors and users, allowing the sensors and their measurements to be accessible and controllable through the Web (Sheth et al., 2008). The SWE suite, although initially designed for sensors, can easily be applied to any type of spatio-temporal data flow (including heterogeneous types of smart devices with an observation capability). It offers a set of specifications in an open standard schema using extensible markup language (XML) and web services. It enables (i) finding sensors and sensor data; (ii) describing sensor systems and data; (iii) recovering real-time and historical sensor observations; (iv) adding simulations and recovering simulation results; (v) reporting results and alerts; and (vi) full web control.

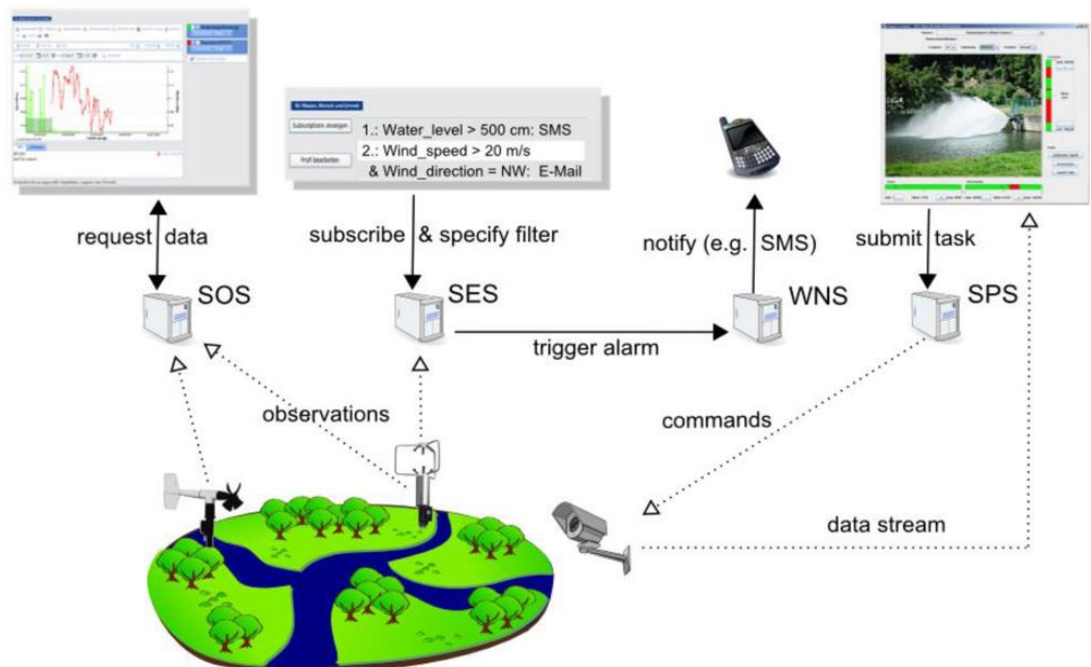

Figure 5. The Sensor Web Enablement suite of standards. Source: Bröring et al. (2011)

SWE (depicted in Figure 5) is organized through several interdependent standards that include the Sensor Model Language (SensorML) (Botts and Robin, 2007), Observations and Measurements (O&M) (Cox, 2007), Sensor Observation Service (SOS), Transducer Markup Language (TransducerML, deprecated) (Havens, 2007), Sensor Planning Service (SPS) (Simonis, 2007), Sensor Alert Service (SAS) (Simonis, 2006) and Sensor Event Service (SES) (Echterhoff and Everding, 2008). In this work, only the first three specifications are shown in detail (i.e. SensorML, O&M, SOS), as they are the most widely used in the IoT context today.

SensorML provides the ability to define a sensor in a structured manner. The standard specifies how to find, process and record sensor observations so that a data model and XML schema can be established to control sensors through the Web. SensorML defines a standard schema describing any type of sensor, stationary or dynamic, in situ or remote, active or passive. The PUCK protocol (O'Reilly, 2010) is an addition to the SensorML standard that provides a low-level protocol to retrieve sensor drivers, and metadata documents, encoded according to SensorML.

The O&M standard, initially developed by the OGC, is also adopted as an International Organization for Standardization (ISO) standard (ISO, 2011). It provides a model for representing and exchanging sensor observations. The standard is encoded using an XML/JSON data model, which describes the relationship between different aspects of the data capture process. The O&M schema defines both observations and phenomena. In addition, it can be extended to better support metadata.

Finally, the SOS provides an interoperable means for serving observations via a Web interface and is the primary service model of the SWE suite. The current version of the standard introduces a modular structure. The base module provides three mandatory operations. The first, "GetCapabilities", offers a spatial and temporal description of the observations that have been stored, as well as a list of the sensors and their available features. The "DescribeSensor" operation is used to return a sensor description using SensorML. The "GetObservation" operation provides access to the actual spatio-temporal data encoded in accordance with the O&M standard.

All the standards described above were conceptualized and adopted several years ago within a completely different technological landscape. The rapid growth of the IoT and the emergence of new technologies (e.g. remote sensing, 4G/5G communication, machine-to-machine and machine-to-human interactions) brought new challenges such as (i) the need for lightweight data encoding, (ii) the need for higher bandwidth for data exchange, and (iii) the issue of constrained devices with little or no computational capabilities, such as RFID tags and QR codes (Kotsev et al., 2018). These challenges acted as a driver for the OGC and led to adoption of new standards that better fit the IoT.

The SensorThings API (Liang et al., 2016), designed to follow the paradigm of the Web of Things (WoT) (Guinard et al. 2010), offers access to data through standard web protocols and is based on the O&M conceptual data model. The main features of the standard are (i) a RESTful interface, (ii) the use of lightweight and efficient JSON encoding, (iii) adoption of the OASIS OData URL pattern (OData) and query options, and (iv) support for the ISO message queuing telemetry transport (MQTT) messaging protocol to offer real-time connections.

The SensorThings API data model (shown in Figure 6) is divided into two parts (profiles), namely, the 'Sensing' profile and the 'Tasking' profile. The former enables IoT devices and applications to CREATE, READ, UPDATE, and DELETE (through the standard web operations HTTP POST, GET, PATCH, and DELETE) IoT data and metadata by invoking a SensorThings API service. In addition, the tasking profile provides a standardized approach for controlling IoT devices through the "ACT" capability, which is revisited in the next section. Each 'Thing' has a Location (or some Historical Locations) in space and time. A collection of Observations grouped by the same Observed Property and Sensor is called a Datastream. An Observation is an event performed by a Sensor that produces a value of an Observed Property of the Feature of Interest.

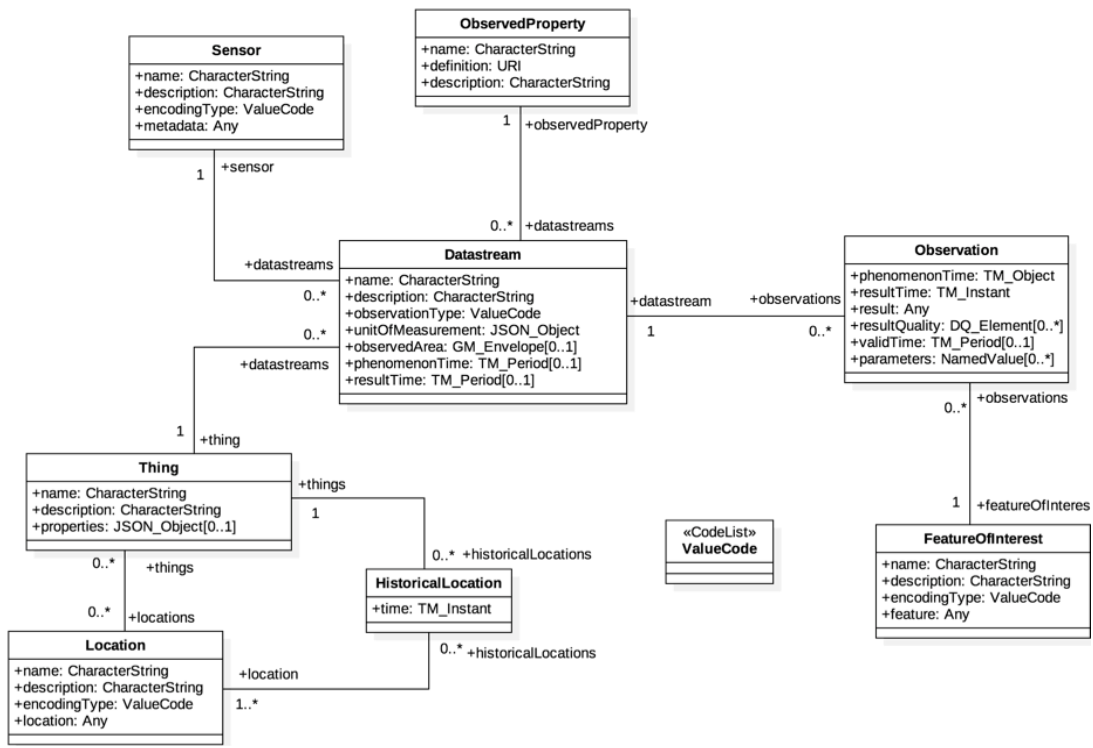

Figure 6. The SensorThings API data model. Each Thing has a Location (or some Historical Locations) in space and time. A collection of Observations grouped by the same Observed Property and Sensor is called a Datastream. An Observation is an event performed by a Sensor that produces a value of an Observed Property of the Feature of Interest. Source: OGC SensorThings API (http://docs.opengeospatial.org/is/15-078r6/15-078r6.html).

From a spatial analysis perspective (Smith et al., 2018), many raster- and vector-based operators and techniques have been developed over the last decades and have been shown to be successful in many varied applications. Substantial progress has been made to bring geospatial workflows -- i.e., a combination of the above spatial operations to accomplish a sophisticated analytical process-- to the cloud and distributed computing environments (e.g., Granell et al., 2010; Granell, 2014; Yue et al., 2016), expanding the field of the Geoprocessing Web (Zhao et al., 2012) to the Digital Earth (Hofer et al., 2018). The OGC Web Processing Service (WPS) (OGC, 2005), a service interface for exposing and executing processes of any granularity on the Web, enables sharing and integration of spatial data processing capabilities on the Web, including polygon area calculation, routing services, or entire environmental models (e.g., Díaz et al., 2008; Granell et al., 2010). The geoprocessing capabilities in DE are extensively covered in other chapters, e.g., Chapter 5, and our interest lies solely in the relationship between the WPS and the IoT (see Section 3.2)

## 3. Interplay between the IoT and DE

One of the aims of this chapter is the identification of potential bridges between the IoT and DE. This overview is partly speculative since we tried to identify potential paths for collaboration between both infrastructures, which may or may not lead to successful linkages in the future. To support our claims in Section 4, we identify the current situation, i.e., the state of the art of the IoT's and DE's technological substrate. In this section, we highlight new technological developments and emerging trends that are or may become crucial in the coming years that were not present or not sufficiently developed at the time of van der Zee & Scholten (2014).

Along the lines of the topics described in Section 2.3, the traditional focus of DE embraces the following high-level functions (Lü et al., 2019): (i) discovery and acquisition of spatial information, (ii) understanding of spatial objects and their relationships (e.g., GIS analysis, spatial statistics), and (iii) determination of the spatio-temporal behavior and simulation rules (e.g., simulations, predictions). These functions help categorize and restrict the discussion in terms of the current technological substrate. However, we should interpret and contextualize these high-level functions of DE from the viewpoint of the IoT.

First, the acquisition of spatial information is a crucial function in the IoT because Things and smart devices observe and sense their environment to collect observational measurements. Through the lens of the IoT, the discoverability of Things and the communication of gathered spatial data become extremely relevant for data acquisition. Of the two main capabilities of Things (see Section 2.2), the ability to **observe and sense**, is a fundamental mechanism to provide input observational data for DE.

Second, spatial statistics and spatial analysis are well-established geospatial methods for exploring spatial patterns, relationships and distributions (de Smith et al., 2018; Worboys & Duckham, 2004). Analytical methods are fundamental building blocks in DE, although recent trends in real-time analysis and edge computing promise to move much of the analytical power to devices (i.e., edge and fog computing) so that gathered data can be immediately processed directly on the smart devices. This trend suggests that analytical improvements in the IoT will also play an important role in DE.

Third, predictive modeling and simulations are required to explore both physical and social dynamic geographic phenomena to better understand the evolution, changes and dynamics of the phenomena from a spatio-temporal perspective, to gain new insights and scientific knowledge to support informed decision-making processes. Understanding spatiotemporal behaviors makes sense from the DE point of view, to aid in the assembly of a detailed yet broad perspective of the complex, multidimensional relationships that occur in the real world. We recognize that prediction and simulation activities are typically associated with DE and that advances in the IoT might contribute to this area, but we see this hypothetical scenario occurring in the mid- to long-term, well beyond the time frame of the speculative exercise in Section 4. Since research on the IoT and DE with respect to predictive modeling and simulations is still in its infancy, we do not cover it in this chapter.

As a result of the previous functions, new scientific knowledge is generated that is necessary for taking informed and insightful actions, often 'acting' over the environment. In terms of **acting**, the second main capability of Things, new knowledge can trigger actions at least at two different levels in the context of the IoT: first, self-calibration of a sensor and/or Thing, similar to adjusting the lens in a human eye to sharpen the image, e.g., changing the sampling frequency; and second, providing a reflex similar to a reaction to pain without thinking, e.g., by opening a valve or level in the case of imminent flooding. However, this view would mean *a priori* that the IoT and Things do not contribute sufficiently to the higher (cognitive) functions of DE such as spatial analysis, predictive modeling and simulation, and the results of higher cognitive functions in DE may impact the acting behavior of Things and the IoT. In addition, we add a fourth function related to the ability of Things to act and take informed actions, depending on the insights and knowledge produced in the analysis, simulations, and predictions in DE.

Figure 7 reflects the existing and potential roles of each infrastructure in relation to the four functions: (i) discoverability, acquisition, and communication of spatial information, (ii) understanding of spatial objects and their relationships, (iii) determining spatio-temporal behavior and simulation rules, and (iv) acting and taking informed actions. We argue that the IoT infrastructure is important in (i) and (iv) whereas DE is more relevant in (ii) and (iii). For (i), the IoT can enhance DE by acquiring data streams from new sources, at a fine scale and high frequency. For (ii), it is plausible that both infrastructures progressively collaborate in a symbiotic manner per use case. From a broader perspective, it can reasonably be argued that DE includes IoT and encompasses the IoT life cycle in a broader ecosystem. Although GIS methods and analysis have traditionally taken a predominant role in DE, the role of the IoT will most likely increase in the future given the close relation between the IoT and the nascent edge-fog-cloud computational paradigms that enable IoT-based analytical processes to be conducted at different scales. This is a partial view, as we focus on the relationship between DE and the IoT. For example, remote-sensing satellite imagery, LIDAR and UAV were intentionally omitted even though they are key spatial data sources (i.e., the first function) for DE. We acknowledge the fuzziness of the boundary between both infrastructures and pay special attention to the interplay between DE and the IoT in Figure 7, demonstrating how collaboration and integration is starting to happen while frictions and barriers are becoming more visible.

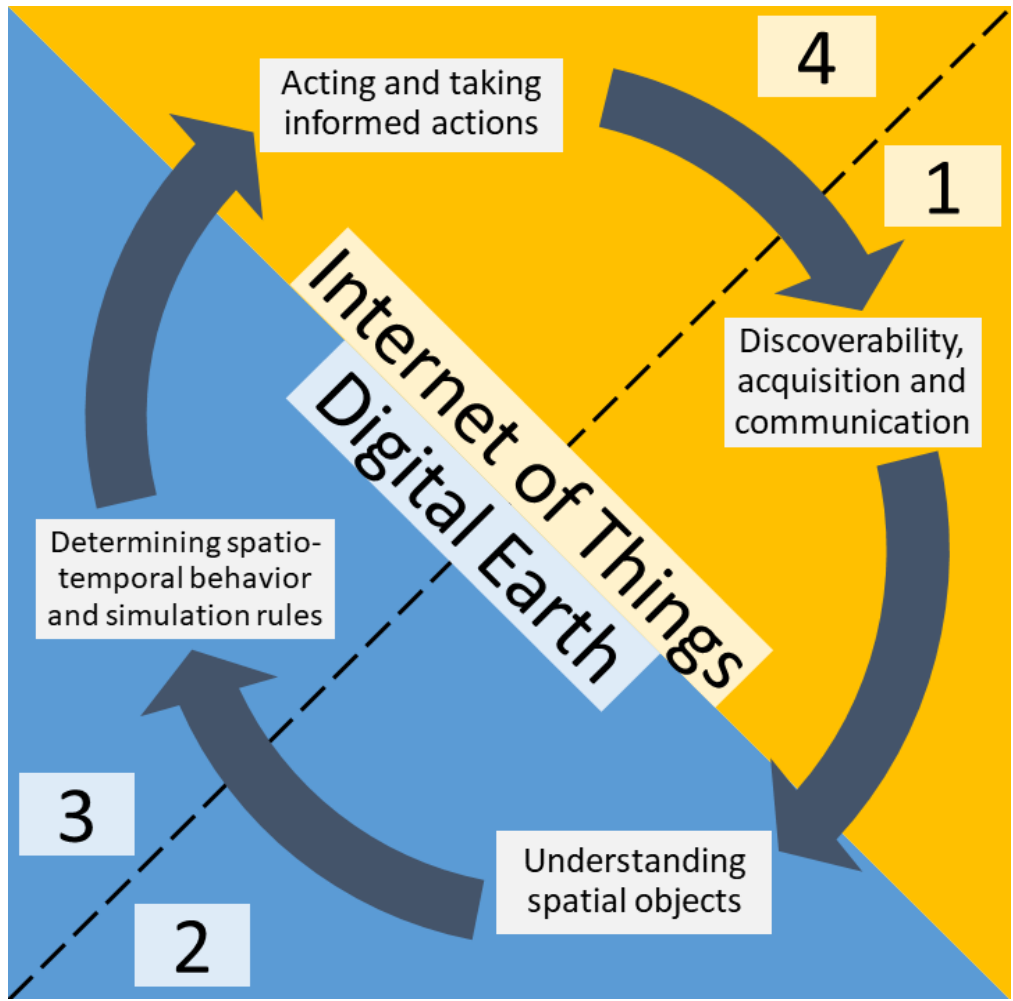

Figure 7: IoT and DE workflow according to the higher cognitive functions in DE.

In the following sections, we identify for all but the third function the current technological substrate.

### *3.1 Discoverability, acquisition and communication of spatial information*

**Discoverability of Things.** An important objective in IoT research is the discovery of devices and their services and/or the data they produce. The absence of standardized discovery methods for the WoT (Zhou et al. 2016) led to the development of online global sensor directories and collections such as Xively (https://xively.com ), SenseWeb (Grosky et al. 2007), SemSOS (Pschorr et al. 2010) and the SWE discovery framework (Jirka, Bröring & Stasch 2009). A key feature of these online directories/registries is that they provide open Web APIs supporting the development of third-party applications. The main drawback is that they are centralized, with a single point of failure. Decentralized approaches have also been proposed, such as IrisNet (Gibbons et al. 2003), which uses a hierarchical architecture for a worldwide sensor Web. G-Sense (Perez, Labrador & Barbeau, 2010) is a peer-to-peer (P2P) system for global sensing and monitoring. These approaches, although more robust and scalable, do not effectively solve the problem of sensor discovery as they still require sensor registration to dedicated gateways and servers, which need to maintain a hierarchical or P2P structure among them.

Approaches towards real-time discovery of physical entities include Snoogle (Wang, Tan, & Li 2008) and Dyser (Elahi et al. 2009). Snoogle is an information retrieval system for WSNs, but it cannot scale for the World Wide Web. Dyser requires an additional Internet infrastructure such as sensor gateways to work. Moreover, utilization of the domain name system (DNS) as a scalable, pervasive, global metadata repository for embedded devices and its extension for supporting location-based discovery of Web-enabled physical entities were proposed (Kamilaris, Papakonstantinou & Pitsillides 2014, Kamilaris & Pitsillides 2012). However, this technique requires changes in the existing Internet infrastructure. It is possible to exploit web crawling for discovery of linked data endpoints, and through them the discovery of WoT devices and services was examined in WOTS2E (Kamilaris, Yumusak & Ali 2016) as well as in SPITFIRE (Pfisterer et al. 2011).

While the approaches described above are mainly targeted at 'professional' users, there is demand for a simple and easy means for the general public to access IoT data. Experts can use a plethora of different service interfaces and tools to discover and utilize data from IoT devices, as implemented by the SmartEmissions platform (Grothe et al. 2016). Nonexpert users typically only search for IoT devices and their data through mainstream search engines such as Google and Bing. Ensuring the discoverability of devices and the data they produce is being investigated for geospatial data in general (see Portele et al., 2016 for further details). A similar approach might be adopted for the IoT, considering its higher complexity due to the high temporal (and spatial) resolution of the data produced by Things.

**Spatial acquisition with Things.** Some examples of geospatial standards to encode sensor metadata and observations were introduced in Section 2.4, and the SensorML standard is one of the most important. SensorML describes sensor metadata in a comprehensive way, providing a useful mechanism to discover sensors and associated observations. This standard specifies information about a sensor such as its sensor operator, tasking services, location, phenomena, and history of the sensor. Thus, it can be used by discovery services to fill their search indexes.

Following the SWE framework, there are two different search types (Jirka et al., 2009): *sensor instance discovery* and *sensor service discovery*. The first type finds individual sensors (devices) or sensor networks, and the second type refers to services that interact with the sensor (through sensing or tasking). Jirka et al. (2009) define three different criteria to identify both annotated search types:

- The Thematic criterion covers the kind of phenomena that a sensor observes, such as temperature, humidity, or rainfall.
- The Spatial criterion refers to the location where the sensor is deployed.
- The Temporal criterion is the time period during which the observations are generated.

This classification was defined from a conventional sensor point of view. The inclusion of current IoT devices with the ability to act leaves the previous criteria incomplete, as some IoT devices act as well as observe. Therefore, the definition of the thematic criterion requires extension to include an IoT device's capability to act, for example, to turn on/off a light or activate/deactivate an air conditioner.

In addition to the three shared criteria, Jirka et al. (2009) defined two criteria that focused exclusively on the *sensor instance discovery* type of search: sensor properties and sensor identification. The sensor properties are based on a specific state of the sensor, for example to find all *online* sensors. The sensor identification refers to the unique id used to identify unambiguously a sensor. Regarding the *sensor service discovery type of search*, two additional criteria were defined: functionality and usage restrictions. The first refers to the functionalities of the associate service such as available operations for data access, alerting or tasking, among others. The second criterion on usage restrictions is related to the permissions and restrictions to access the service functionalities.

Two different aspects are vital for the successful discovery of a sensor: metadata and semantics. As for all spatial data, metadata is essential to describe and discover a sensor or a network of sensors. SensorML was created for this purpose and can define a sensor in a well-known manner to add flexibility and allow for the use of any type of sensor. The sensor instance registry (SIR) defines operations for handling sensor metadata and allows for sensor discovery. The above criteria, both common and specific for each type of search, are closely related to the metadata aspect for the discovery of sensor instances and services.

Semantics is the other pillar in a powerful and effective discovery service. Semantic rules can aid in locating sensors related to the same phenomena or discovery of all sensors that are related to the same thematic aspect. This semantic view can be extrapolated to link sensors with places to retrieve sensors or observations associated with place names. The sensor observable registry (SOR) offers a primary interface to explore this kind of relationship between phenomena and sensors.

Unfortunately, the support of semantics is a weakness in the SWE standards. To solve this issue, an initiative from World Wide Web Consortium (W3C) was created to integrate and align sensors with semantic web technologies and Linked Data. This contribution was led by the W3C Semantic Sensor Network Incubator Group (SSN-XG) that proposed an ontology called Semantic Sensor Network (SSN) to address the semantic gap in sensor-related OGC standards (Compton et al., 2012). The main fields of this ontology are sensors (e.g., location, type), properties (e.g., precision, resolution, and unit), and measurements (values).

Despite the great advances that SSN brought, it does not currently support all the possibilities that the IoT offers since SSN was designed before the mainstream adoption of the IoT. New ontologies have been launched to cover this gap. One example is how the Internet of Things Ontology (IoT-O). IoT-O adds some missing concepts relevant to the IoT such as Thing, Actuator, and Actuation (Seydoux et al. 2016). Similarly, the IoT-O is a follow-up to SSN, the lightweight ontology SOSA (Sensor, Observation, Sample, and Actuator). It is the result of a joint effort of the W3C and OGC that builds on the lessons learned from SSN to provide a better representation of the IoT and alignment with OGC-related specifications (Janowicz et al. 2018).

**Communication with Things.** The advances in IoT connectivity solutions such as Bluetooth, ZigBee, Wi-Fi and 3-5G (Palatella et al., 2016) combined with decreases in the price and energy consumption of IoT components have led to a huge deployment of smart devices using IP-connectivity worldwide, increasing the frequency of communication to the point that they are perceived as always connected. As outlined above, these devices can offer two different capabilities, observing (sensing) and acting. A decade ago, sensor networks were only able to capture and send data, similar to a simple data logger. In recent years, the ability to establish two-way communication between Things and the cloud has added the feature that Things can (re)act. Consequently, new protocols that enable machine-to-machine (M2M) communication have been developed, with the goal of providing efficient and transparent two-way communication channels between smart devices. Examples of such TCP/IP-based protocols are the advanced message queuing protocol (AMQP), MQTT, and the simple/streaming text oriented messaging protocol (STOMP). These communication protocols are adapted to the requirements of IoT devices that are constrained concerning their performance and energy efficiency.

*3.2 Spatial understanding of objects and their relationships*

**Spatial analysis of Things.** There are many more smart devices (Things) around today than five years ago. Smart devices now produce massive volumes of data, i.e., flows of data with strong temporal and spatial features. Therefore, spatial analytical methods such as proximity, area, volume, and trajectory are of vital importance in analyzing processes of Things. However, the variety of data sources related to the IoT has posed new analytical challenges, especially in the design and provision of a new class of analytical tools capable of handling real-time temporally and spatially referenced data from a plethora of heterogeneous smart devices (Trilles et al., 2017). Despite the existence of tools capable of analyzing temporal data in real time, the same does not appear to be true for the spatial component. Space (location and orientation for all Things, size and shape for larger Things such as cars) plays an indispensable role in the IoT, as Things-generated data have spatial properties and are spatially related to each other. Promising initiatives and platforms have recently emerged with the aim of performing spatio-temporal analysis in real-time, such as Microsoft Streaminsight, the Oracle Spatial Database with the Oracle Complex Event Processing engine, and the GeoEvent processor module as an extension of the ArcGIS Server environment (ArcGIS Server, n.d.).

Despite these notable efforts, spatial support for the real-time analysis of IoT data is still in its infancy. As van der Zee & Scholten (2014) noted, any IoT architecture should consider the geospatial component. Location provides a kind of 'glue' that efficiently connects smart devices. The authors proposed storing the location of each 'Thing' and other geographic-related features such as orientation, size, and shape. However, the ability to handle and analyze the location of Things in near real time is still limited with existing analytical platforms, despite its opportunities (McCullough et al., 2011; Rodríguez-Pupo et al., 2017).

Furthermore, spatio-temporally located Things have the potential to significantly improve advanced geospatial analysis, as Kamilaris and Ostermann (2018) describe in their review on the potential role of geospatial analysis in the IoT field. In short, Kamilaris and Ostermann suggest network analysis and monitoring, surface interpolation, and data mining and clustering as spatial analysis techniques and methods that would especially benefit from an increasing number of mobile or stationary sensor Things. However, as the authors noted, these advanced analytical applications have been scarcely exploited to date.

**Geospatial standards for Things.** Despite some remarkable exceptions such as prototype systems to analyze data from air quality sensor networks (Trilles et al., 2015b), real-time, geospatial analysis approaches and tools have not been sufficiently developed to offer standardized procedures through uniform interfaces that can be widely consumed and integrated in DE applications. DE has traditionally considered sensors as a fundamental pillar to collect information to support and realize strategies or policies at a higher level. As described in Section 2, the SWE suite was the initial step in offering a standardized specification that would fulfil the requirements demanded by the IoT from the DE perspective. For example, the SOS specification requires handling large XML documents, which is problematic in a typical scenario in the IoT where memory capacity and connectivity are limiting factors.

Although the core of the SWE suite has served to cover the required functionality of the IoT, the complexity of the data models in some of the specifications (Tamayo et al., 2011, Trilles et al., 2014) and the appearance of new requirements such as the ability to work in real time and to act have reduced the applicability and integration of the SWE suite in the scope of the IoT. In an effort to bridge the gaps between SOS and the IoT, new extensions or approaches attempt to make the SOS interfaces more suitable for IoT devices. These approaches include SOSLite (Pradilla et al., 2015), TinySOS (Jazayeri et al. 2012) and SOS over CoAP (Pradilla et al. 2016).

Another crucial feature for the analysis functionality of the IoT and Things is the ability to specify and perform real-time and asynchronous notifications and communications. In this regard, the GeoMQTT protocol based on the MQTT protocol allows for adding spatial notification and data streaming between publish/subscribe instances (Herle et al., 2018). Following the original approach of the MQTT channels, the authors proposed the concept of GeoPipes to distribute instances and enable the sharing of geospatial data streams in a standardized manner.

Laska et al. (2018) proposed a real-time stream processing pipeline that allows for spatiotemporal data stream integration from IoT devices. A data integration layer allows for geospatial subscriptions using the GeoMQTT. Tools such as Apache Kafka and Storm are used to transfer and apply map matching algorithms to IoT data with spatiotemporal components. For example, these algorithms were used to analyze traffic congestion for a recent route optimization using IoT Things with global navigation satellite system (GNSS) receivers in buses.

Another study (Rieke et al., 2018) took an additional step to bridge the DE and IoT realms by arguing for the need to establish event-driven architectures as a natural evolution of the predominantly static spatial data infrastructures (SDI). The authors identify a series of interdependent issues that need to be addressed in the coming years to take full advantage of the uptake of eventing in GIScience (and DE). The issues relate to the (i) inconsistencies between classic data access methods that are based on a request-response pattern, and event-driven approaches where a publish-subscribe pattern prevails, (ii) heterogeneous approaches for defining event patterns, (iii) multiple standards and limited support in software tools, (iv) the integration of devices in an SDI and the data they produce, and (v) the lack of semantic interoperability of geospatial events.

### *3.3. Taking informed actions and acting over the environment (ACT)*

As shown in the defined IoT lifecycle (Figure 7), to act means to take or perform actions (over the environment) depending on the results obtained in previous functions. Bélissent (2010) noted that this feature can make the management of public services in a city, education, health, safety, mobility or disaster management more aware, interactive and efficient.

IoT devices have been traditionally suitable for use as input sources for decision support systems (DSSs) in a multitude of application domains and use case scenarios such as disaster management, cities, mobility, and safety. In this chapter, we focus on spatial decision support systems (SDSSs), which are defined as interactive systems designed to support decision making related with spatial planning problems. SDSSs have evolved to more complex architectures and communication models, from systems deployed on the cloud operating with data from the WSN (or IoT data sources) to a shift in the computing paradigm in which the actual computation is implemented at three different levels: edge, fog, and cloud (Figure 8). In this new setting, both the computation and decisions are made closer to the producers of the data (*Things*).

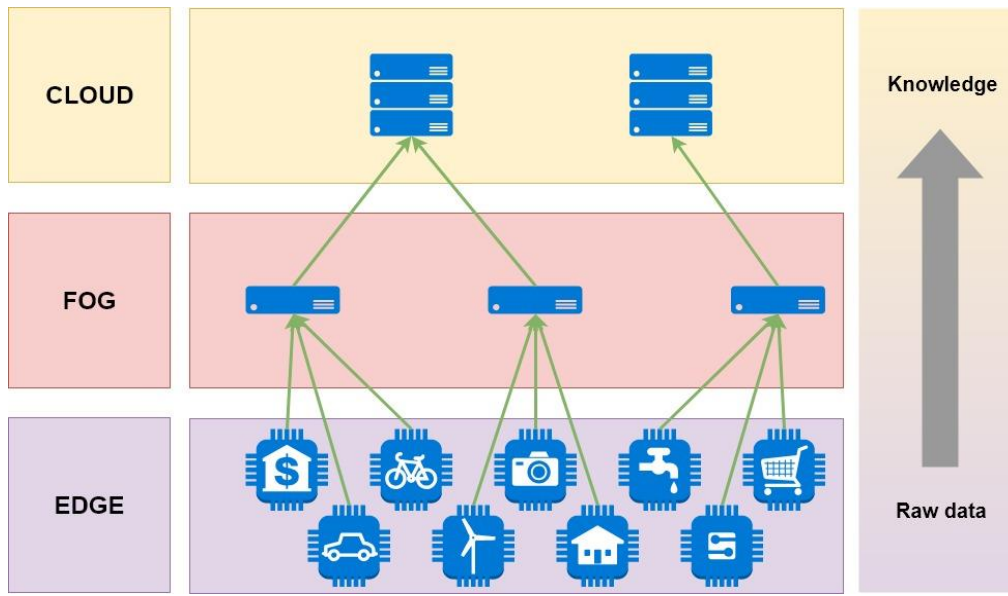

Figure 8: Three-layer IoT architecture

The 'Edge' is the layer that covers the smart devices and their users, providing local computing capacity within Things. The 'Fog' layer is hierarchical, aggregating a variable number of edge layers. In addition to computing, the fog layer has other functionalities such as networking, storage, control, and data processing, possibly using data produced by the edge layer and data from other sources. As a result, data contextualization is more important in the fog layer to make sense of different data sources than the typical single data stream in an edge layer. The 'Cloud' layer on top performs the final analysis to extract information and create knowledge to be transferred for decision support actions. This implies an increased level of contextualization and complexity in the analysis process than in the previous (lower) layers, at the cost of losing capacity for real-time analysis.

Given the edge-fog-cloud layered architecture, the introduction of geospatial concepts and spatial analysis in the fog layer could allow for decision-making processes without a human in the loop based entirely on the semantics of the spatial-temporal dimensions in the incoming data. In recent years, many efforts have been made to move the analysis from the cloud to the fog layer, with the aim of reducing latency in the analysis once the data are received in the fog layer (Barik et al. 2016).

Although data usually flow from the edge to the cloud layer (sensing capability), devices with the ability to act (tasking) also require information to perform their operations. The tasking capability allows for other devices or users to actuate devices via the Internet so that these 'controlling' devices or users can easily control them to execute tasks remotely. Autonomous Things would be previously programmed to act without establishing a connection. While the sensing capability allows for users to continuously monitor the status of devices and the environmental properties they capture, the tasking capability can help users make adjustments accordingly by controlling devices remotely.

In general, combining the sensing and tasking capabilities of IoT devices enables users to create various automatic and efficient tasks and applications. These kinds of applications are called "physical mashup" applications (Guinard et al. 2010). A simple, domestic example is the activation of an air conditioning system depending on the position and behavior of the user, through an application that uses a GNSS sensor. In this example, the air conditioning device provides an interface to turn on/off (tasking) the system to establish a comfortable temperature. To facilitate this kind of mashup of sensing and tasking capabilities, a uniform (interoperable) interface for users or applications to enable access and communication is a critical requirement.

The tasking feature was initially conceived in the SPS specification of the SWE suite. SPS offers a standardized interface for tasking sensors and sensor systems and defines interfaces to expose sensor observations and metadata. For example, a sensor network can be set up to measure air pollution in 5-minute intervals or a satellite can be tasked to remotely sense a specific region on the surface of the globe (De Longueville et al. 2010). This standard offers operations such as GetFeasibility, which can be used in advance to verify whether the execution of a task is feasible for a certain sensor, and the DescribeResultAccess operation to determine the access points to collected data. The SPS interface also offers functionality for managing submitted tasks, including convenient operations for retrieving the status of a task, updating tasks or cancelling them.

A next step is the tasking profile of the SensorThing API, which is a follow-up, improved profile of the SPS (Simonis, 2007). The SensorThing API (see Section 2) defines two different profiles, Sensing and Tasking. The Tasking profile is based on the SPS standard and enables interoperable submission of tasks to control sensors and actuators. The main difference between SPS and the SensorThings API is that the former offers task operations over sensors and the latter also includes tasks on actuators. Although the first version of the SensorThing API did not include the Tasking profile, a new candidate standard illustrates the potential of the SPS standard, duly adopted and aligned with the requirements of the SensorThings specification (Liang & Khalafbeigi, 2018). This new specification called Tasking Core defines three new entities, *TaskCapability*, *Task,* and *Actuator* (Figure 9).

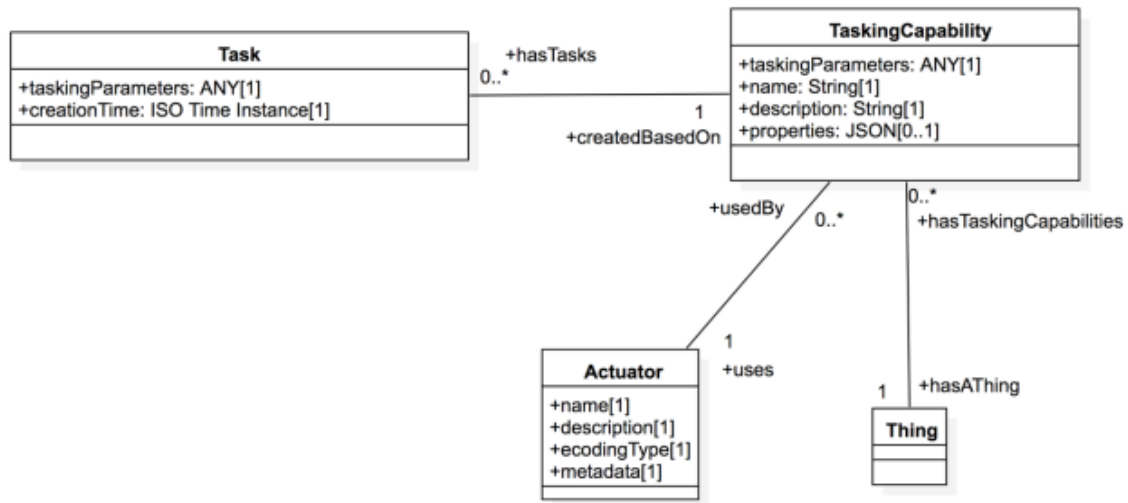

Figure 9: The SensorThings API Tasking Entities. Source: OGC SensorThings API (http://docs.opengeospatial.org/is/15-078r6/15-078r6.html)

The *TaskingCapability* entity describes all supported tasks for each Thing and how they can be used. This entity is defined by four properties: name, description, taskingParameters, and properties. The second entity, *Task*, is a list of performed tasks that are defined by a set of tasking parameters (commands executed) and creation time. The last entity is the *Actuator* and defines a type of transducer that converts a signal to a real-world action or phenomenon. This entity is comprises a name, description, encoding type of metadata and metadata.

**4. Case studies on smart scenarios**

In this section, we show how the IoT and DE work hand-in-hand in real-world scenarios based on the latest technology initiatives to relate the IoT and DE described in the previous section. Kamilaris & Ostermann (2018) provide an extensive overview of work at the nexus of geospatial analysis and the Internet of Things; here, we provide a selection of case studies in various domain applications, with a special focus on the relationship between DE and the IoT.

In the context of applications for environmental monitoring and resource management in cities, recent examples of IoT applications include an Arduino-based sensor platform in Seoul to measure variations in the physical-chemical parameters in water streams (Jo & Baloch, 2017). The sensor platform is powered by solar energy and transmitted sensor readings every second via Bluetooth for three years. Although the case study in Jo & Baloch (2017) relies on a single sensor station and the clustering analysis of the raw data focuses uniquely on the temporal dimension, the paper shows the potential of Arduino-based sensing modules for environmental sensing applications in smart city applications. To improve solid waste management, Tao & Xiang (2010) developed an information platform to support recycling. The main technologies were RFID and GPS to track and check waste flows between collection, transport, and processing facilities. Lee et al.

(2015) examined the role of the IoT in an industrial service provision scenario (fleet management) and Fazio & Puliafito (2015) use the example of road conditions to showcase a cloud-based architecture for sensor and data discovery. They distinguish two scenarios of data- or device-driven search, and develop the system architecture based on the OGC SWE suite and the extensible messaging and presence protocol (XMPP).

Reducing the required energy consumption remains an important objective for IoT devices. Ayele et al. (2018) proposed a dual radio approach for wildlife monitoring systems. They combine Bluetooth low energy for intraherd monitoring with LoRa for low-power wide-area networks to communicate between herd clusters and a monitoring server. The proposed architecture promises significant advantages in reducing power consumption while maintaining low latency.

Improving traffic management is another promising IoT application area. In 2006, Lee et al. proposed the use of cars as a mobile vehicular sensor network and for data exchange in "smart mobs". More recently envisioned solutions include parking management and smart traffic lights as part of a cognitive road management system that handles different types of traffic efficiently (Miz & Hahanov 2014). Jing et al. (2018) examined the combination of GNSS localization and RFID tagging for infrastructure asset management with promising results. Additionally, the city of Aarhus in Denmark deployed traffic sensors across major roads in the city, and the information was used by the CityPulse project to provide context-aware recommendations to users for route planning (Piu et al. 2016).

Noise pollution is a frequent problem in dense urban areas, and because urban morphology makes noise distribution modeling difficult, it has attracted participatory sensing approaches. Wireless acoustic sensor networks are another option. Segura Garcia et al. (2016) presented a case study in the small city of Algemesi (Spain), where a network of 78 inexpensive sensor nodes based on Raspberry PIs collected sufficient data for a subsequent highly accurate spatial interpolation.

Okasenen et al. (2015) harnessed movement data from mobile sports tracking applications in urban areas to produce heat maps of cyclists commuting through the city of Helsinki. Mobile phones could be considered IoT sensor devices in participatory sensing-based models for mining spatial information of urban emergency events, as demonstrated by Xu et al. (2016). In addition, van Setten (2004) supported the COMPASS tourist mobile application with context-aware recommendations and route planning. Mobile phones were also used for crowdsourcing-based disaster relief during the Haitian earthquake (Zook, Shelton & Gorman 2012), where people used the camera and GPS of their phones to send information from the field to the authorities to map the landscape of the disaster and assess the overall damage.

University campuses present an interesting environment for smart city approaches because the visitors are usually more tech-savvy than the average population, the network coverage is good, and the geographic boundaries allow for a comparatively crisp delineation of the study area. Cecchinel et al. (2014) presented a system architecture for a smart campus case where the four requirements of sensor heterogeneity, reconfiguration capability, scalability, and data as a service were handled via a middleware in the Amazon Web Services (AWS) cloud, with Arduino Uno and Raspberry Pi sensors for bridging. Another case study at a university campus examined the impact of nearby weather and pollution sensors on the everyday decision-making of the students (Kamilaris & Pitsillides 2014). Trilles et al. (2015a) presented a sensorized platform proposal that adheres to the principles of the IoT and the WoT. They use the SensorThings API to avoid interoperability issues. An environmental WSN in a Smart Campus scenario was developed as a proof of concept.

However, smart approaches with IoT technology are not limited to smart city applications. Savant et al. (2014) presented a low-cost automated weather station system for agriculture that uses Raspberry Pi systems at its core and SWE to transmit data. The sensor readings were also broadcast on a dedicated Twitter account. The system has been extended with additional components such as a web-based client (Savant et al. 2017). The environmental impact of agriculture was studied by Kamilaris et al. (2018) in the region of Catalonia, Spain. In their study, sensors measuring nitrates and data from the mobile phones of farmers in the region were used. Fang et al. (2014) presented a holistic approach to environmental monitoring and management through an integrated information system that collects data on the regional climate for the city of Xinjiang from various sources including IoT sensors, and related it with ecological response variables such as the primary production and leaf area index. For environmental monitoring, the AirSensEUR project established an affordable open software/hardware multisensor platform, which can monitor air pollution at low concentration levels to create maps of pollution levels in different areas (Kotsev et al. 2016).

A crucial component of any DE system and application is monitoring shifting surface conditions such as erosion on sandy beaches. Pozzebon et al. (2018) presented an Arduino-based system to measure the height of sandy beaches and dunes in real-time. The sensor network uses the ZigBee standard to transmit data, with a GPRS transmitter for sending sensor readings to a MySQL database. Another example is the monitoring of landslides in mountainous areas. Benoit et al. (2012) tested a successful cheap wireless sensor network using XBee for communication and GPS for localization. A thematically related case study is the use of small and inexpensive sensors for monitoring and early-warning systems for floods caused by melting snow in the Quergou River basin (China), as reported by Fang et al. (2015). In addition, changing climate conditions make reliable and efficient management of storm water surges in urban areas important. Rettig et al. (2016) designed and tested a geospatial sensor network for this task, built using common, off-the-shelf components.

With respect to the provision and reception of cultural heritage and cultural services, Chianese et al. (2017) proposed and tested a system that combines business intelligence, Big Data, and IoT data collection to analyze visitor interests and behaviors in a museum.

Although IoT devices were only part of the approach, measuring visitor proximity to artworks, their integrated use with other technologies and platforms showcases the strength of a multisensory DE approach.

**5. Frictions and synergies between the IoT and DE**

Based on the current technological substrate that provides the initial steps to establish connections between the IoT and DE according to the three cognitive functions (Section 3), and the presentation of selected case studies (Section 4), in this section, we (i) carry out a speculative exercise to discuss the main existing limitations and frictions that prevent the IoT and DE from working closer together and (ii) suggest future ways to establish effective communication channels between the two infrastructures.

Before going into detail, it is necessary to establish a fundamental assumption that influences any discussion related to the frictions and synergies between the IoT and DE: the diverging speeds of development of DE and the IoT. New technology and disruptive breakthroughs generally challenge the status quo in any sector, and adopting such improvements can enable more rapid developments and new applications. However, the rapid growth of the IoT field has produced a vast variety of IoT devices and protocols and, consequently, the landscape of IoT-related standards, protocols and specifications is fragmented. For example, a large portion of 'Things' were not originally designed to connect to the Internet; they were later adapted to establish Internet connections by adding connectivity chips via microcontrollers (e.g., Arduino, Raspberry Pi) or through tags (QR Code or RFID). As a result, many different ways to connect hardware and software to enable Internet connectivity were developed and established with no clearly agreed upon consensus and consequently resulted in a lack of interoperability. This example illustrates the great variety and complexity of the IoT universe, where the exponential growth of the IoT is due to the rapid decrease in the size, cost, and energy requirements of sensors, and the ubiquity of network coverage for wireless Internet connections, leading to many standardization efforts following diverging paths. In addition, DE has been traditionally characterized by a slow adaptation of new improvements (López 2011), and thus, the recent technological developments **have not evolved at the same speed in DE as in the IoT.** Noting this fundamental friction, we identify other potential frictions and synergies, which may be considered two sides of one coin, and organize the discussion according to the cognitive functions defined in Section 3.

*5.1 Discoverability, acquisition and communication of spatial information*

A direct result of the fragmented standardization context noted above is the **absence of well-accepted global protocols for the discovery of Things,** which also occurs to some extent in DE. Search and discovery is crucial for geo-locating nearby, local, and/or relevant real-world devices and services, a vital step in exploiting sensor data and services to create more advanced knowledge. Early efforts in this direction are discussed in Section 3.1, but we are still far from a complete solution to this difficult problem, which must be addressed along with the challenges of better description of devices and services and the semantics of the data involved, especially from a geospatial point of view.

Therefore, it remains an open issue to build an IoT-DE ecosystem in a way that will be compatible with standardized IoT reference models and architectures to enable the discovery of relevant sensors (or Things) and related services. Although there are many different scenarios and solutions, several common features can be extracted to find synergies between both infrastructures: the modularity and interoperability of IoT components, open models and architectures, flexible service compositions, integrated security solutions, and semantic data integration. There is an intensified effort regarding the development of architectural frameworks and solutions such as the IEEE or ITU-T models, as well as other related works and approaches developed under the auspices of IETF, W3C, or OASIS. From a DE point of view, associated services for sensor devices and instances are the cornerstone to enable seamless communication and interoperability between the IoT and DE. There are different options such as the SWE and SensorThings API, the latter of which is especially relevant for the establishment of potential solid bridges between the IoT and DE concerning common data models for better data acquisition and unified interfaces for enhanced sensor and service discovery. Some research works have already made substantial progress. Jara et al. (2014) presented a comprehensive framework and architecture to enable discovery over a wide range of technologies and protocols, including legacy systems, and Wang et al. (2015) implemented annotations with an ontology-based semantic service model, SPARQL queries, and geographic indexing to enable sensor discovery in an experimental study, which delivered faster and more accurate responses than other tested approaches.

*5.2. Spatial understanding of objects and their relationships*

A friction between DE and the IoT is related to the way geographical features are modeled. Traditional GIS data models conceptually abstract the real-world objects into core geometric elements such as points, lines, polygons, and volumes, implemented as raster data models, vector data models, or a combination. These data models were designed to perform spatial analyses such as distance computations and topological operations. Despite these great achievements, GIS (and DE) data models were not designed to cope with the richness and complexity of the interactions between the physical, natural, and social actors that naturally occur in the environment in the way that the IoT potentially can. As noted above, smart devices and Things can 'sense' the environment in a way that was unimaginable before, and, consequently, the streams of rich and finer data acquired by **IoT devices do not fit well with the "coarse-grained" vector/raster data models** widely used in DE applications and systems, as these spatial structures were not intended to handle data with such a high spatio-temporal resolution.

The lack of suitable data models to efficiently manage data at high spatio-temporal resolution highlights **the need for new tools to process data coming from Things and smart devices** in which the modeling of geospatial features has not yet been fully resolved. Moreover, real-time data is often a defining feature in the IoT, as IoT devices and Things can produce data at a high frequency (e.g., data streams), which requires methods for real-time analysis. Therefore, the lack of new algorithms and implementations for real-time computation and processing streams of spatially referenced data sets is a clear limitation. Although some tools can run geospatial queries of stored data, they do not offer ways to analyze data from IoT devices and sensor nodes in real-time (Nittel, 2015).

Unlike the IoT, any changes in the DE arena have been more gradual and less frenetic. However, some notable changes indicate the way forward to consolidate potential bridges between DE and the IoT in the midterm and long term. For example, in a Digital Earth Nervous System (De Longueville et al. 2010), **Things could perform basic geospatial operations on sub-networks of Things**, providing processed information for the higher-level elements of a DE. Geometric measurements and basic geospatial analysis are application areas in which Things have been used more widely in recent years (Kamilaris and Ostermann, 2018). Similarly, an often overlooked component of IoT applications are the gateway nodes that connect the sensor devices to the wider network. In addition to a simple routing function, these gateways can perform other tasks including exploratory analysis (clustering, event-detection) of incoming data. Rahmani et al. (2018) examined the use of smart gateways in an e-health system that monitors several individual physiological parameters, demonstrating the potential benefits of (spatial) analysis executed directly on smart gateways in the context of DE-related applications such as precision agriculture, environmental monitoring, and disaster management.

The status quo of services for spatial analysis and geoprocessing on the Web is mainly driven by the WPS standard specification (Section 2.4). However, Herle & Blankenbach (2018) argued that the current WPS standard is not well suited to handle the large amounts of real-time streaming data expected from massive IoT sensor networks. Building on previous work, they extended the WPS with the GeoPipes concept using the GeoMQTT protocol for communication, implementing several smaller proofs-of-concept for application cases such as inverse distance weighting with a sliding window and trajectory data mining. In addition, Armstrong et al. (2018) presented an IoT+CyberGIS system to detect radiation risk and propose that new approaches are needed to integrate the IoT and geospatial analysis and support the fourth scientific paradigm of data-intensive discovery (Hey et al., 2009).

*5.3. Taking informed actions and acting over the environment*

In the initial stages of DE, it was thought that sensors could only capture what is happening in the physical environment, i.e., sensors as mere data loggers. The data collected by these sensors are transferred from bottom to top until reaching the SDI repositories. In this sense, the IoT is much more complex because, in addition the feature of acting on the physical environment, the IoT supports communication between devices in the same layer (edge) and complex strategies to determine solutions to real, large problems can be developed. As mentioned above, DE should be adapted to the possibilities that the IoT devices can offer to enrich the capabilities of the current SDIs.

The previously noted heterogeneity problem of connecting IoT devices implies different hardware specifications across the multiple IoT devices. This variety of hardware means that the abovementioned standards cannot work at a low level. This is why the standards mainly define web service interfaces, and connectors or adapters (hub approach) are required to control IoT nodes. Similar to the hub approach, the sensor interface descriptor (SID) solution is a declarative model based on the Sensor ML standard for describing device capabilities (Broering & Below, 2010), sensor metadata, sensor commands, and device protocols. In terms of the tasking capability, the SID describes device protocols with the open systems interconnection (OSI) model using an XML schema and thus understanding and adapting the SID may be costly for IoT device manufacturers.

An opportunity that DE can offer the IoT is a global vision on the in situ data that the IoT collects, with the aim of establishing strategies to perform actions in a coordinated manner among the IoT nodes, taking advantage of the ability to act. To conclude, the following table (Table 1) summarizes the frictions and synergies between the IoT and DE.

Table 1. Detected frictions and synergies between the IoT and DE.

|  | **Discoverability, acquisition and communication of spatial information** | **Understanding spatial objects and their relationships** | **Taking informed actions and acting over the environment** |
|---|---|---|---|
| Frictions | -Absence of well-accepted global protocols for the discovery of Things | -IoT devices do not fit well with coarser vector/raster data models<br>-Lack of tools to process data from Things | -DE has traditionally considered sensors as collectors, with data flowing from bottom to top.<br>-GIS standards must be adapted for each hardware specification |
| Synergies | -Different standardized IoT models and architectures such as SWE and SensorThings API | -Things can perform basic geospatial operations<br>- Some initiatives have adapted GIS processing standards to support IoT data | -DE provides a global view to establish IoT node strategies to act |

**6. Conclusion and outlook for the future of the IoT in support of DE**

The concept of combining sensors organized in networks to monitor the environment has been around for decades, and DE has contributed to its expansion. The confluence of new technologies has created a new reality that offers millions of new possibilities, led by the IoT revolution that promises to create a newly interconnected "smart" world (or Earth). After the massive deployment of a ubiquitous array of IoT devices and the impact it made, the world cannot give up being 'online'. Today, the IoT has enabled millions of relationships between objects and Things, so that objects, people, and their environment are more tightly intertwined than ever. Despite the great advances achieved in recent years, like all disruptive innovations, the IoT presents a series of challenges that should be treated as a priority in the coming years, especially in the areas of security, interoperability and standards, privacy, and legal issues. DE can also play a crucial role in handling some of these challenges.

The IoT and DE dichotomy presents various challenges that should be addressed in the near future to create a more beneficial union for both parties: The first challenge is to activate mechanisms to streamline the adaptation of new IoT functionalities from DE. Traditionally, DE is characterized by its comparative inertia to adopt new approaches that imply improvements in terms of performance or usability. Examples include the slow adoption of more flexible interfaces such as the RESTful web interface or data formats that are more suitable for exchange such as JSON in sensor standards such as the SOS specification (Tamayo et al., 2011). The tradeoffs between standardization and disruptive innovation in DE should be carefully discussed by all involved actors to fuel rapid, innovative developments in DE like those in the IoT field. Although the standardization process is key to establishing permanent links between the two infrastructures, it should not slow down innovative changes and technical developments, and standards should be seen as a means to filter out and embrace changes that prove to be useful, effective and valuable for improvement of the IoT-DE ecosystem.

When a technological field grows exponentially, it often leads to heterogeneity and variety in the short term. Within the IoT, this is partly due to the impact that the continuous development and improvement of hardware technology has on IoT devices. Therefore, another challenge to be addressed is the heterogeneity of IoT devices. Although the OGC specifications have helped in the service connection and data/service access levels, the IoT still presents a wide variety of different hardware developments and implementations, most of which are disconnected from the DE infrastructure, and therefore remain invisible for DE applications. The development of ad hoc adapters is one way, at least until a standards consensus is reached in the IoT field, to allow for interaction with the variety of hardware specifications of IoT devices and Things and foster connections between the two infrastructures. This is not an optimal solution since the integration of IoT devices is a challenging and difficult task, but it helps discern the connections and adaptors that may eventually become candidates for standardization bodies.

Throughout this chapter, we revisited many tools that are capable of analyzing spatially referenced data collected by IoT devices. However, the quantity and quality of tools that handle the temporal dimension of data in real time far exceeds those that deal with the spatial dimension. An additional barrier is the large-scale variance in the data models between IoT devices and the decision-making systems that are typically established in DE. Optimal spatial models to handle scale variations can be useful to analyze the information received from IoT devices and obtain a more high-level vision that can be interpreted by decision makers and policy makers. Therefore, investment in the research and development of better tools to spatially analyze IoT data in real time on the edge, fog and cloud scales is a priority in the IoT-DE ecosystem roadmap.